\documentclass[preprint]{aastex}

\def\ltsima{$\; \buildrel < \over \sim \;$}
\def\simlt{\lower.5ex\hbox{\ltsima}}    
\def\gtsima{$\; \buildrel > \over \sim \;$}
\def\simgt{\lower.5ex\hbox{\gtsima}}    

\def\ref{\par\noindent\hangindent 20 pt}

\def\mincir{\ \raise -2.truept\hbox{\rlap{\hbox{$\sim$}}\raise5.truept 
\hbox{$<$}\ }}  %
\def\magcir{\ \raise -2.truept\hbox{\rlap{\hbox{$\sim$}}\raise5.truept %

\hbox{$>$}\ }}

\received{}
\accepted{}
\journalid{}{}
\articleid{}{}
\shortauthors{Falomo et al.}
\shorttitle{Host galaxies of high z quasars}
\slugcomment{Version \today}

\input psfig.sty

\begin{document}

\title{\bf Near-infrared imaging of the host galaxies of three radio--loud
quasars at z $\sim$ 1.5}

\author{Renato Falomo}
\authoraddr{Osservatorio Astronomico di Padova, Vicolo dell'Osservatorio 5,
35122 Padova, Italy}
\authoremail{falomo@pd.astro.it}
\affil{Osservatorio Astronomico di Padova}

\author{Jari Kotilainen}
\authoraddr{Tuorla Observatory, University of Turku, V\"ais\"al\"antie 20,
FIN--21500 Piikki\"o}
\authoremail{jarkot@deneb.astro.utu.fi}
\affil{Tuorla Observatory, Finland}

\author{Aldo Treves}
\authoraddr{Universit\`a dell'Insubria, Como, Italy}
\authoremail{treves@astmiu.uni.mi.astro.it}
\affil{Universit\`a dell'Insubria, Como, Italy}

\begin{abstract}

We present high spatial resolution near--infrared $H$-band (1.65 $\mu$m)
images, taken with ISAAC on UT1 of ESO VLT, of three radio-loud quasars at
z$\sim$1.5, as a pilot study for imaging of a larger sample of radio-loud and
radio-quiet quasars in the redshift range 1 $<$ z $<$ 2. We are able to
clearly detect the host galaxy in two quasars (PKS 0000-177 and PKS 0348-120)
and marginally in the third (PKS 0402-362). The host galaxies appear compact
(average bulge scale-length R(e) $\sim$ 4 kpc) and luminous (average
M(H) = --27.6$\pm$0.1). They are $\sim$2.5 mag more luminous than the typical
galaxy luminosity  (M*(H) = --25.0$\pm$0.2), and are comparable to the
hosts of low redshift radio-loud quasars (M(H) $\sim$--26), taking into
account passive stellar evolution. Their luminosities are also similar to
those of high redshift radio galaxies.
All three quasars have at least one close companion galaxy at a projected
distance $<$ 50 kpc from the quasar, assuming they are at the same redshift.

\end{abstract}

\keywords{Galaxies:active -- Infrared:galaxies -- Quasars:general}

\section{Introduction}

During the last decade, a wealth of information was obtained on the
close environment and, in particular, on the host galaxies of quasars, using
both Hubble Space Telescope (HST, e.g. Disney et al. 1995;
Bahcall et al. 1997; Boyce et al. 1998; McLure et al. 1999) and ground-based
4-m class telescopes (e.g. McLeod \& Rieke 1994, 1995; Taylor et al. 1996;
Kotilainen, Falomo \& Scarpa 1998; Kotilainen \& Falomo 2000;
Percival et al. 2000). Significant improvements in spatial resolution and
instrumental efficiency have allowed the characterization of the properties
(mainly luminosity and scale-length) of quasar hosts and, at least at low
redshift (z $<$ 0.5), to investigate in detail their morphology.  At low
redshift (and to a lesser extent at intermediate redshift, 0.5 $<$ z $<$ 1),
it is now well established that the galaxies hosting these powerful AGN are
predominantly ellipticals or early-type spirals. Radio-loud quasars (RLQ) are
almost exclusively found in galaxies dominated by the bulge (spheroidal)
component. These luminous ellipticals exceed by $\sim$2--3 mag the typical
galaxy luminosity  (M$^*_H \sim$ --25;
Mobasher, Sharples \& Ellis 1993), and are close  to the
brightest cluster galaxies (M$_H \sim$ --26; Thuan \& Puschell 1989). For
radio-quiet quasars (RQQ), the situation is less clear, with both types of
hosts found. 

Our present knowledge of quasar host galaxies is essentially limited
to z $<$ 1 (see references above). This has enabled only a preliminary
insight into the cosmological (z--dependent) quasar--host galaxy
connection. This evolution should become much clearer in the redshift
interval 1 $<$ z $<$ 2, since z$\sim$2 is close to the epoch of the
most vigorous nuclear activity when the host galaxies are still
young. The observed reasonable similarity of the cosmic quasar
evolution with the rate of galaxy formation (e.g. Franceshini et al
1999) may represent the overall effect of a fundamental link between
massive galaxies and their nuclei, that has driven their formation
history.   Deep high
spatial resolution HST images of distant galaxies (e.g.  Abraham et
al. 1996; Koo et al. 1996; Le Fevre et al. 2000) have begun to trace
the galaxy formation, while very little is still known about the
evolution of distant quasar hosts. This is due to the very rapid
increase with redshift in the difficulty of detecting quasar hosts,
because of the (1+z)$^{4}$ cosmological dimming of the surface
brightness. Consequently, only a few optical or near-infrared (NIR)
studies of RLQ hosts at z $>$ 1 have been conducted so far, either
using conventional techniques (e.g.  Lehnert et al. 1992, 1999;
Aretxaga et al. 1998) or, more recently, with adaptive optics
(e.g. Hutchings et al. 1999), and suggest for RLQs extremely luminous
hosts with very high star formation rates ($\geq$ 2000 M$\odot$
yr$^{-1}$) that have no counterparts in the local universe
(e.g. Tresse \& Maddox 1998: Treyer et al 1998) or in high redshift
field galaxies (e.g. Steidel et al. 1996; Lowenthal et al. 1997). On
the other hand, similar studies of RQQs at z$\sim$2.5 (Lowenthal et
al. 1995) were unable to resolve the hosts, suggesting they are at
least 1 mag fainter than the RLQ hosts, and indicating that the two
types of quasars have distinct types of hosts, rather than being
different only in their level of radio emission.

The study of high redshift quasar host galaxies and their immediate
environment is difficult, because of the often overwhelming brightness of the
nucleus with respect to the host that becomes progressively fainter at a few
arcsec from the nucleus. The two main requirements are, therefore, good
spatial resolution (narrow point spread function: PSF) and large collecting
area of the telescope for the detection of faint diffuse features (e.g. the
host galaxy and close companions).

We present here the results of a pilot study of NIR imaging of three
RLQs (PKS 0000-177, PKS 0348-120 and PKS 0402-362) at z
$\sim$1.5 extracted from a list of quasars 
with redshift 1.0 $<$ z $<$ 2.0, -25.5 $< M_V
<$ -28 and --60$^\circ$ $< \delta <$ -8$^\circ$ and with 
sufficiently  bright stars
within the field of view of the quasar for the characterization of the
PSF. Hubble constant H$_0$ = 50 km s$^{-1}$ Mpc$^{-1}$ and
deceleration parameter q$_0$ = 0 are used throughout this paper.

\section{ Observations and data analysis}

NIR $H$-band (1.65 $\mu$m) images of the quasars were obtained at the
European Southern Observatory (ESO) in Paranal, Chile, using the ISAAC camera
(Cuby 2000) mounted on the first 8m unit telescope (UT1) of the Very Large
Telescope (VLT). The SW arm of ISAAC is equipped with a 1024 x 1024 px Hawaii
Rockwell array, with pixel scale 0\farcs147 px$^{-1}$, giving a field of view
150 x 150 arcsec (corresponding to $\sim$ 1.8 Mpc at z = 1.5). The
observations were performed in service mode during the nights of 20 and 21
October 1999. Each quasar was observed for a total integration time of 1 hour
using a jitter procedure and individual exposures of 2 minutes per frame. The
jittered observations were controlled by an automatic template (see
Cuby 2000), which produced a set of frames slightly offset in telescope
position from the starting point. The observed positions were randomly
generated within a box of 10 x 10 arcsec centered on the first pointing. The
seeing was very good during the observations, resulting in stellar FWHM of
0\farcs50, 0\farcs41 and 0\farcs38 in the fields of PKS 0000-177,
PKS 0348-120 and PKS 0402-362, respectively.

Each frame was flat-fielded and sky-subtracted and the final image was
produced for each quasar by co-adding these frames. Data reduction was
performed by the ESO pipeline for jitter imaging data (Devillard 2000). The
normalized flat field was obtained by subtracting ON and OFF images of the
illuminated dome, after interpolating over bad pixels. Sky subtraction was
done by median averaging sky frames from 10 frames nearest in time.
The reduced frames were aligned to sub-pixel accuracy using a fast
object detection algorithm, and co-added after removing any spurious pixel
values. Photometric calibration was performed
against standard stars observed during the same night. The estimated internal
photometric accuracy is $\pm$0.03 mag.

In order to determine the amount of extended emission around the quasars, one
needs to perform a detailed study of the PSF. The relatively large field of
view of ISAAC ($\sim$ 2\farcm5) and the constraint on the selection of the
quasars to have at least one sufficiently bright field star, allowed us to
perform a reliable characterization of the PSF. For each field, we analyzed
the shape of many stellar profiles and constructed a composite PSF, whose
brightness profile extends down to $\mu_H$ = 24.5 mag arcsec$^{-2}$. This
warrants a reliable comparison between the luminosity profile of the quasars
and of the stars without requiring extrapolation of the PSF. The shape of the
PSF profile was found to be stable across the field of the images.

For each quasar, we have derived the azimuthally averaged fluxes excluding
all regions around the quasars contaminated by companion objects. This
procedure yielded for each quasar the radial luminosity profile out to a
radius where the signal became indistinguishable from the background noise.
For our observations, this level corresponds to $\mu$(H)
$\sim$24  mag arcsec$^{-2}$, typically reached at $\sim$2$''$ distance from
the nucleus. Straightforward comparison of this profile with the scaled PSF
gives us a first indication of the amount of the extended emission. Detailed
modeling of the luminosity profile was then carried out using an iterative
least-squares fit to the observed profile, assuming a combination of a point
source (PSF) and an elliptical galaxy described by de Vaucouleurs law,
convolved with the proper PSF.

We also attempted a fit using a pure exponential disk model for the host
galaxy. Note that the small extent of the hosts and the dominance of the
nuclear emission made it impossible to discriminate between the two models.
Consistently with the properties of lower redshift RLQs, we have assumed the
elliptical model for the determination of the host galaxy luminosities. If a
disk model is assumed, the luminosities  become
$\sim$0.2 -- 0.3 mag fainter. This difference does not affect the main
conclusions of this study.

The basic properties and the derived parameters of the quasars are given in
Table 1, where column (1) gives the name of the quasar, column (2) the
redshift, columns (3) -- (4) the $H$-band apparent magnitude of the nucleus
and the host, columns (5) -- (6) the $H$-band absolute magnitude of the
nucleus and the host, and column (7) the scale-length of the host. Absolute
magnitudes have been K-corrected (at this $z$ K-coor is $<$ 0.2 mag in H) ) using the optical-NIR evolutionary synthesis
model for elliptical galaxies (Poggianti 1997). 

\section{ Results}

In Fig. 1, we show the final images of the three quasars together with their
PSF-subtracted images. The PSF scaling factor was
adjusted in order to prevent strong negative values in the few central pixels
of the residual image. Although this method tends to over-subtract the point
source, it gives us a model-independent way to analyze the surrounding
nebulosity. After the subtraction of the scaled PSF, two out of the three
quasars images exhibit clearly extended emission and suggest the presence of
knotty structure. 
For one quasar (PKS 0402-362), only marginal extended emission is detected.
Detailed results for individual quasars are reported below. The uncertainty on the host galaxy luminosity depends mainly on the relative emission and size of the nebulosity compared with that from the nucleus. In order to estimate the error on the host magnitude we 
fitted the profiles with variable values of host and derived the level at which the 
fit was  unacceptable ($\Delta\chi_\nu^2 >$ 3). Based on this 
procedure, we estimate the uncertainty of the host galaxy
magnitudes to be $\sim$ 0.3 mag ($\sim$0.5 mag for PKS 0402-362), while the
uncertainty on the scale-length may be as high as 50\%.

\subsection{PKS 0000-177}

This quasar is located at $\sim$4$''$ NW from an m$_H$ $\sim$14.5 mag star
which, although well separated from the quasar, has been subtracted from the
final image to avoid confusion. In our $H$-band image (Fig. 1a), PKS 0000-177
at z = 1.465 appears round and compact with a companion object at $\sim$2$''$
SW. After removing this companion, we note faint ``arc-like'' structures to
the NE and SW at $\sim$1\farcs5 from the nucleus, which, however, are not well visible in the  image reproduction (see Fig 1a). 
After subtraction of a scaled PSF (Fig. 1b),
extended emission is clearly detected. This nebulosity exhibits also a faint
knot at $\sim$ 0\farcs6 E of the nucleus. This knot may be related to the
double radio structure at 5 GHz, separated by $\sim$0\farcs7 along E-W
direction (Aldcroft et al. 1993). After masking this companion
we derived the brightness profile (Fig. 2a) that exhibits a significant
excess over the scaled PSF, starting at $\sim$1$''$ (12 kpc) radius from the
nucleus. Modeling of this profile shows that the extended emission
corresponds to a host galaxy with M$_H$ = --27.4 and R(e) $\sim$ 3.5 kpc. 
A companion galaxy (m$_H$ = 20.3) is visible at 1\farcs8 distance SW of the
quasar. If it is at the redshift of the quasar, its projected distance is
$\sim$21 kpc and its absolute magnitude is M$_H$ = --25.8.

\subsection{PKS 0348-120}

PKS 0348-120 is a flat spectrum radio quasar at z = 1.520 (Wright et
al. 1983). NIR photometry by Wright et al., J = 17.7 and K = 16.1, is
consistent with our H = 17.1. The $H$-band image of the quasar
(Fig. 1c) is clearly extended and elongated approximately along NE-SW
direction.  A resolved companion galaxy (G1) with m$_H$ = 20.5 lies at
2\farcs2 NE from the quasar. The projected distance of this galaxy, if
at the redshift of the quasar, is 27 kpc. Another, fainter galaxy (G2)
with m$_H$ = 21.1 is situated at 3\farcs2 (39 kpc) SW, while a much
fainter extended emission (G3; m$_H$ = 22.2) is visible at 3\farcs9
(43 kpc) W of the quasar. 
Modeling of the radial brightness profile (Fig. 2b) confirms that this
quasar is well resolved and its host galaxy is very luminous (M$_H$ =
--27.6) and compact (R(e) = 4.9 kpc). The profile is not smooth,
suggesting the presence of substructure in the host. Straightforward
subtraction of a scaled PSF (Fig. 1d) reveals a knot at 0\farcs7 (8.6
kpc) from the nucleus along the direction of G1. This feature is
better emphasized in the deconvolved image of the quasar (see inset in
Fig. 1c) using a Lucy-Richardson algorithm implemented in IRAF (Lucy
1991 ) with 15 iterations. With the present data, it is not possible
to clarify the nature of this circumnuclear emission knot. Most
likely, it is another small companion galaxy projected onto the
envelope of the quasar host, but higher spatial resolution images
(either from space or from ground with adaptive optic) will be
necessary to reveal its morphological details.

\subsection{PKS 0402-362}

Although PKS 0402-362 is at about the same redshift (z = 1.417) as the other
quasars in this study, its nucleus is $\sim$2 mag brighter. Our photometry
(H = 15.1$\pm$0.1) is comparable with H = 14.9$\pm$0.2 measured by
Peterson et al. (1976). The $H$-band image of this quasar is shown in
Fig. 1e. The nuclear luminosity is very high, M(H) = --30.6, and it out-shines
the light from the host galaxy. After subtraction of the scaled PSF
(Fig. 1f), no significant extended emission is detected. Using azimuthally
averaged radial brightness profile, we substantially improved the
signal-to-noise 
and found a small excess of light over the PSF starting from 1$''$ distance
from the nucleus (Fig. 2c). Modeling of this profile, assuming it is due to an elliptical
galaxy, yields a host  luminosity of M(H) = --27.7 and
scale-length R(e) $\sim$ 4 kpc. Comparison of  various  stellar
profiles in the field indicates that this light excess is real. However,
since it is small, we consider this only a marginal host detection. The
uncertainty of the host galaxy luminosity is $\sim$0.5 mag. 
We detect a companion galaxy with m$_H$ = 19.5 at 2\farcs8 SW from the
quasar. Assuming it is at the same redshift as the quasar, its projected
distance is 34 kpc, and it appears to be very luminous (M$_H$ = -26.6; only 1
mag fainter than the quasar host). Note, however, that the optical spectrum
of the quasar exhibits a weak absorption system tentatively identified as
MgII 2800 \AA ~at z = 0.797 (Surdej \& Swings 1981) which, if associated with
the companion galaxy, would lead to a reduced M(H) = -24.5.

\section{Discussion}

We have presented deep, high spatial resolution (FWHM $<$ 0\farcs5) NIR
images of three high redshift RLQs. In all three cases, we are able to
resolve the quasar, although for PKS 0402-362 only marginally. The detected
extended emission is most naturally interpreted as due to starlight from the
galaxies hosting the quasars. The derived NIR host luminosities
(M$_H \sim$ --27.5) are very bright and correspond to galaxies $>$ 2 mag
brighter than M$^*$ (M$_H$ = --25; Mobasher et al. 1993). The nuclei are
$\sim$1.5 mag ($\sim$3 mag for PKS 0402-362) more luminous than their host
galaxies. The scale-length of the hosts ranges from 3 to 5 kpc and appear to
be smaller than those observed in low (z $<$ 0.5) and intermediate
(0.5 $<$ z $<$ 1) redshift quasar hosts (e.g. McLeod \& Rieke 1994;
Taylor et al. 1996; McLure et al. 1999; Kotilainen et al. 1998;
Kotilainen \& Falomo 2000), which typically have scale-lengths $\sim$10 kpc.
This is similar behavior to what is found in early-type galaxies in the HDF-N
(e.g. Fasano et al. 1998).

The luminosities of the host galaxies of our three RLQs are very
similar to those of high redshift 6C radio galaxies (RG) studied by
Eales et al. (1997).  In their Hubble diagram, RGs at z $\sim$1.5 have
$K$-band magnitude $\sim$18.5, while we find $<$ m$_K$(host) $>$
$\sim$18.3 (assuming H--K = 0.2). The similarity with high $z$ RGs is
also supported by the knotty structure observed in the hosts of two of
the quasars and in the density of the immediate environment (cf
e.g. Best Longair and R\"ottgering 1997).  
Assuming a standard passive stellar evolution of the host galaxies, we
expect that in the $H$-band luminous ellipticals at z $\sim$1.5 should
be $\sim$1.4 mag brighter than ellipticals at the present epoch
(Poggianti 1997). The hosts of these quasars therefore would
correspond to still very luminous (M$_H$ $\sim$ --26) early-type
galaxies at z = 0. Similarly high luminosities have been found for RLQ
hosts at redshifts z $<$ 0.5 (e.g.  McLeod \& Rieke 1994; Taylor et
al. 1996; McLure et al. 1999) and at 0.5$ < z < $1 (M(H) $\sim$--27;
Kotilainen Falomo \& Scarpa 1998; Kotilainen \& Falomo 2000).

These results underline a passive evolutionary scenario for RLQ and RG
hosts that appears significantly different from that of RQQ hosts.  In
fact a recent study of RQQs at z $\sim$1.8 and z $\sim$2.7, performed
with HST and NICMOS in the $H$-band, was able to resolve 4 out of 5
RQQs (Ridgway et al. 2000). The deduced host luminosities appear 1-2
mag fainter than those found in our RLQ hosts and in the 6C RGs (Eales
et al. 1997), and are similar to M*. This suggests that the systematic
difference of host luminosity between RLQs and RQQs, already noted at
low redshift (e.g. Bahcall et al. 1997 and references therein), is
even more significant at higher redshift, possibly indicating a
different formation and/or evolutionary history of the two types of
AGN depending on whether or not they can develop radio emission.

Models of galaxy formation and evolution based on hierarchical
clustering (e.g. Kauffmann \& H\"ahnelt 2000) predict progressively
less luminous host galaxies for quasars at high redshift. This seems
to be in agreement with the observations of high $z$ RQQ hosts
(Ridgway et al. 2000; Hutchings 1995), which may still be undergoing
major mergers to evolve into the low redshift giant ellipticals, or do
not significantly evolve to become low $z$ M* galaxies. This contrasts
with our results on high $z$ RLQs and with previous indications that
RLQ hosts at $z$ = 2-3 are extremely luminous (Lehnert et al. 1992)
and have no counterpart in the local Universe. On the other hand,
similarly to our results, Ridgway et al. (2000) found that the hosts
of RQQs are compact (scale-length $\sim$ 4 kpc) and show high
frequency of close companions.

The combined comparison of the host galaxy properties of RLQs in the
redshift interval 0.1 $<$ z $<$ 1.5 thus suggests a scenario where
luminous AGN live in very luminous massive galaxies whose star
formation rate and evolution appear indistinguishable from those
expected for a (normal) non-active elliptical galaxy. 

The large-scale environments of the three quasars observed do not appear
particularly rich in galaxies with respect to the background. There is no
evidence of a rich cluster of galaxies around any of the quasars. However, in
all three cases we have detected a companion galaxy of m$_H$ $\sim$ 20$\pm$1
at a distance from the quasar $<$ 3$''$ (40 kpc). Based on the counts of
46 resolved objects detected in the images we derive an average density of
galaxies with 19$< m(H) <$ 23 (corresponding to m$^*$-2 $< m< $ m$^*$+1 at
z $\sim$1.5), in the 1 Mpc projected area around the quasar, of 11 galaxies
arcmin$^{-2}$. We thus expect to find a chance projection of $\sim$0.3
companions in the three fields within 3$''$ from the quasar. The Poissonian
probability of finding 3 such companions is P = 0.003.  This indicates that
the observed companions are likely related to the quasars. 

Close companion galaxies around quasars have been noted since the
first systematic imaging studies of quasars (e.g. Hutchings \& Neff
1990; Hutchings 1995; Bahcall et al. 1997; Hutchings et al. 1999) and
spectroscopic investigations have shown that, at least in the case of
the closest companions, they are often galaxies at the same redshift
of the quasars (e.g. Heckman et al. 1984; Ellingson et al. 1994;
Canalizo \& Stockton 1997).  For example, in the HST study of 20
nearby quasars, Bahcall et al. (1997) found a high frequency of close
($<$ 25 kpc) companions. Their intrinsic luminosities (assuming the
redshift of the quasar) ranges from M$_V$ $\sim$ --18.5 to M$_V$
$\sim$ --22.5. If the companions observed around the three quasars in
this study are at the redshift of the quasars, their luminosities are
quite high (M$_H$ $\sim$ --24.5 -- --26.5). This is similar to
the results at low redshift (Bahcall et al.) if a modest amount
($\sim$1.3 mag) of galaxy evolution is taken into account and a
typical (V -- H $\sim$ 3.5) galaxy color is used.  Somewhat less
luminous companion galaxies seem to occur around BL Lac objects
(Falomo et al 2000) but a larger statistic is needed to draw a sound
conclusion on this point.

{\bf\noindent Acknowledgments.}\\
This work was partly supported by the Italian Ministry for University and
Research (MURST) under grant Cofin98-02-32 and Cofin 98-02-15. This research
has made use of the NASA/IPAC Extragalactic Database (NED), which is operated
by the Jet Propulsion Laboratory, California Institute of Technology, under
contract with the National Aeronautics and Space Administration.\\

{\bf\noindent References}\\

\noindent

Abraham, R.G. van den Bergh, S. Glazebrook, K., Ellis R.S., Santiago, B.X., Surma, P., Griffiths, R.E. 1996, ApJS 107, 1\\
Aldcroft, T., Elvis,M., Bechtold,J. 1993, AJ 105, 2054\\
Aretxaga, I., Terlevich, R.J., Boyle,B.J., 1998, MNRAS 296, 643\\
Bahcall, J.N., Kirhakos, S., Saxe, D.H., Schneider, D.P., 1997, ApJ 479, 642\\
Best, P.N., Longair, M.S. and R\"ottgering, H.J.A 1997 MNRAS 292, 758\\
Boyce, P.J., Disney, M.J., Blades, J.C. et al., 1998, MNRAS 298, 121\\
Canalizo, G., Stockton,A., 1997, ApJ 480, L5\\
Cuby, J.G., 2000, ISAAC user manual, ESO\\
Devillard, N., 2000, Eclipse User's guide, ESO internal report\\
Disney, M.J., Boyce, P.J., Blades, J.C., et al., 1995, Nat 376, 150\\
Eales, S., Rawlings,S., Law-Green, D., Cotter, G., Lacy, M., 1997, MNRAS 291, 593\\
Ellingson, E., Yee, H.K.C., Bechtold, J., Dobrzycki, A., 1994, AJ 107, 1219\\
Falomo R., Scarpa R.,  Treves A. and Urry C.M. 2000 ApJ in press (astro-ph/0006388)\\
Fasano G., Cristiani S., Arnouts S., Filippi M. 1998 AJ 115, 1400\\
Franceshini A., Hasinger G., Miyaji T, Malquori D. 1999, MNRAS 310, 5\\
Heckman, T.M., Bothun, G.D., Balick,B., Smith, E.P. 1984, AJ, 89, 958 \\
Hutchings, J.B., 1995, AJ, 109, 928 \\
Hutchings, J.B., Neff S.G. 1990, AJ, 99, 1715 \\
Hutchings, J.B., Crampton, D., Morris, S.L., Durand, D., Steinbring, E., 1999, AJ 117, 1109\\
Kauffmann G., H\"ahnelt M., 2000, MNRAS 311, 576\\
Koo, D., Vogt, N.P., Philips, A.C., Guzman, R., Wu, K.L. et al. 1996, ApJ 469, 535\\
Kotilainen J.K., Falomo R., 2000, A\&A submitted\\
Kotilainen  J.K., Falomo R., Scarpa R., 1998, A\&A 332, 503\\
Le Fevre, O., Abraham, R., Lilly, S.J., Ellis, R.S., Brinchmann, J. et al. 2000, MNRAS 311, 565\\
Lehnert, M.D., Heckman, T.M., Chambers, K.C., Miley, G.K., 1992, ApJ 393, 68\\
Lehnert, M.D., van Breugel, W.J.M., Heckman, T.M., Miley, G.K., 1999, ApJS 124, 11\\
Lowenthal, J.D., Heckman, T.M., Lehnert,M.D., Elias, J.H., 1995, ApJ 439, 588\\
Lowenthal, J.D., Koo, D.C., Guzman, R. et al., 1997, ApJ 481, 673\\
Lucy,L., 1991, ST-ECF Newsletter 16, 6\\
McLeod, K.K., Rieke, G.H., 1994, ApJ 431, 137\\
McLeod, K.K., Rieke, G.H., 1995, ApJ 454, L77\\
McLure, R.J., Kukula, M.J., Dunlop ,J.S. et al. 1999, MNRAS 308, 377\\
Mobasher, B., Sharples, R.M., Ellis, R.S., 1993, MNRAS 263, 560\\
Peterson,B.A., Jauncey, D.L., Wright,  A.E., Condon, J.J. 1976, ApJ 207, L5\\
Percival, W.J., Miller, L., McLure, R.J., Dunlop, J.S., 2000, MNRAS, in press\\
Poggianti, B.M., 1997, A\&AS 122, 399\\
Ridgway,S., Heckman,T., Calzetti,D., Lehnert,M. 1999, Lifecycles of Radio Galaxies (eds. J.Biretta et al.), New Astronomy Reviews (astro-ph/9911049)\\
Steidel, C.C., Giavalisco,M., Dickinson,M., Adelberger,K.L., 1996, AJ 112, 352\\
Surdej, J., Swings,J.P., 1981, A\&AS 46, 305\\
Taylor, G.L., Dunlop, J.S., Hughes, D.H., Robson, E.I., 1996, MNRAS 283, 930\\
Thuan, T.X., Puschell,J.J., 1989, ApJ 346, 34\\
Tresse, L., Maddox,S.J., 1998, ApJ 495, 691\\
Treyer,M.A., Ellis, R.S., Milliard, B., Donas, J., Bridges, T.J. 1998, MNRAS 300, 303\\
Wright,A., Ables,J.G., Allen,D.A. 1983, MNRAS 205, 793\\

\clearpage
\begin{table}
\begin{center}
\begin{tabular}{llllllll}
\multicolumn{8}{c}{Table 1 - Properties of the quasar host galaxies}\\
Quasar   &  z &  m$_H$(nuc) & m$_H$(host) & M$_H$(nuc) & M$_H$(host)& R$_e$(kpc)\\
\hline \\
 (1)     & (2)    & (3)   &   (4)    & (5)   & (6)   & (7) & \\

PKS 0000-177  &  1.465 &  16.8 &  18.7 & -28.9 & -27.4 & 3.6    \\
PKS 0348-120  &  1.520 &  17.1 &  18.6 & -28.9 & -27.6 & 4.9    \\
PKS 0402-362  &  1.417 &  15.1 &  18.2 & -30.6 & -27.7 & 4.1    \\
\hline \\
\end{tabular}
\end{center}
\end{table}

\clearpage

\figcaption[vltqso_fig1.ps]{$H$-band images of the three quasars
before (top) and after (below) subtraction of a scaled PSF (see
text). From left to right: {\it a),b)} PKS 0000-177, {\it c),d)} PKS
0348-120 and {\it e),f)} PKS 0402-362. The full size of the images in each
panel is $\sim$ 17$''$.  North is to the top and east to the left.
The inset in panel {\it c)} shows the result of a deconvolution of the
image. The inset in the panel {\it d)} yields a different grey-scale
of the central portion of the PSF subtracted image in order to enhance
the knot structure. }

\figcaption[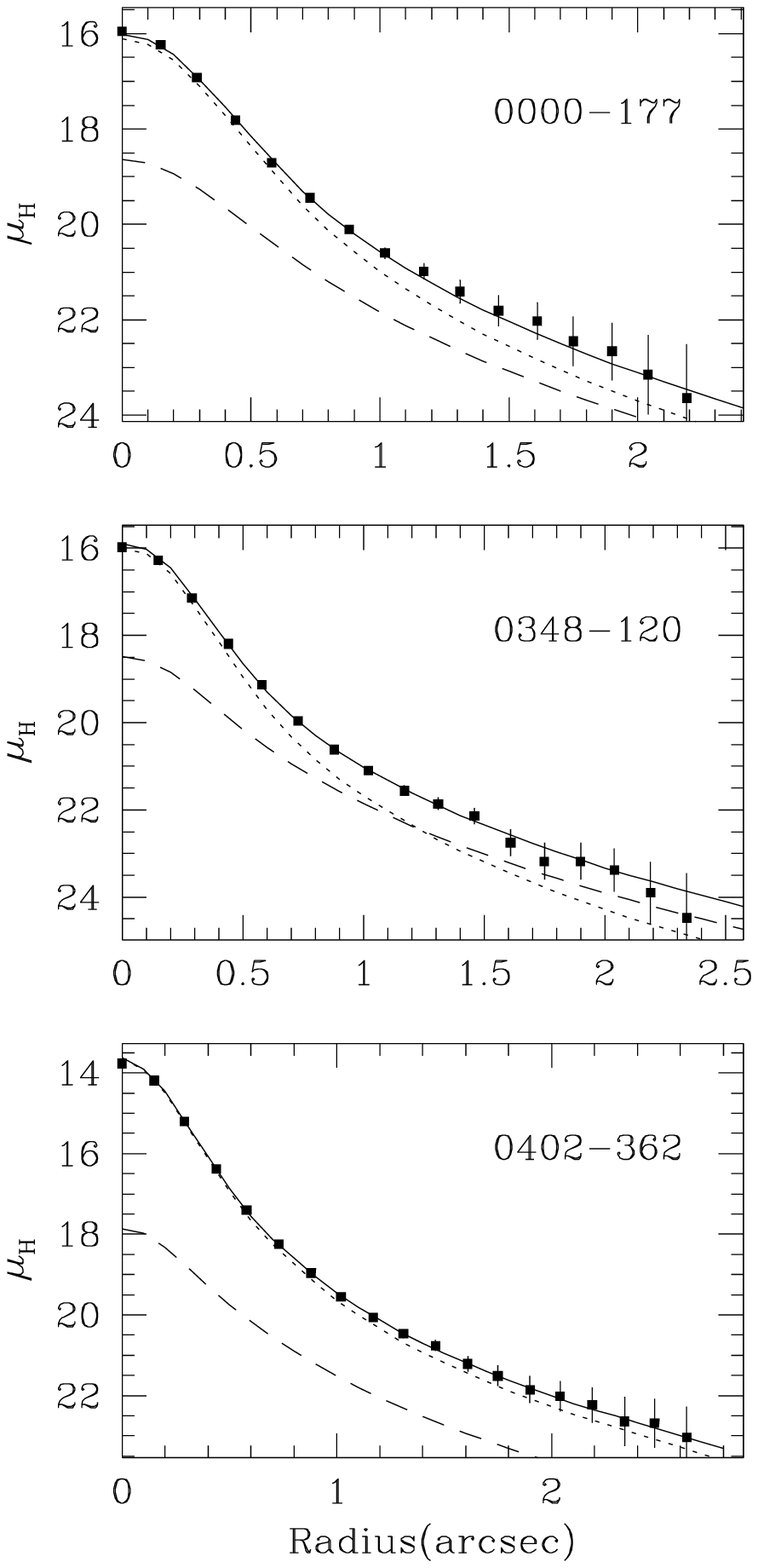]{The observed radial brightness profiles in
the H band of the three quasars (filled squares), superimposed to the
fitted model consisting of the PSF (short-dashed line) and the de
Vaucouleurs bulge (long-dashed line). The solid line shows the total
model fit.}

\end{document}